\documentclass[aps,prl,reprint,floatfix]{revtex4-1}

\usepackage[pdftex]{graphicx}
\DeclareGraphicsExtensions{.pdf,.png}

\usepackage{grffile}

\usepackage{hyperref}
\hypersetup
{
pdfmenubar=true, 
pdfhighlight=/O, 
pdfpagelayout=SinglePage, 
pdffitwindow=true, 
colorlinks=true, 
linkcolor=blue, 
citecolor=blue, 
urlcolor=blue 
}

\usepackage{amsmath, amsthm, amssymb, amsfonts, amsbsy}
\usepackage{mathtools} 
\usepackage{bm}

\renewcommand*{\v}[1]{\boldsymbol{#1}}
\newcommand\bnabla{\boldsymbol{\nabla}}
\newcommand\bcdot{\boldsymbol{\cdot}}
\renewcommand{\div}{\bnabla\bcdot} 
\newcommand{\m}[1]{\v{#1}}

\renewcommand{\d}{\mathrm{d}} 

\newcommand{\upi}{\pi}

\newcommand{\abs}[1]{\lvert #1 \rvert}
\newcommand{\norm}[1]{\lvert #1 \rvert}

\usepackage[capitalise]{cleveref}
\crefformat{equation}{Eq. (#2#1#3)} 
\Crefformat{equation}{Equation (#2#1#3)} 
\crefrangeformat{equation}{Eqs. (#3#1#4--#5\crefstripprefix{#1}{#2}#6)} 
\Crefrangeformat{equation}{Equations (#3#1#4--#5\crefstripprefix{#1}{#2}#6)} 
\crefmultiformat{equation}{Eqs. (#2#1#3)}{ and~(#2#1#3)}{, (#2#1#3)}{ and~(#2#1#3)} 
\Crefmultiformat{equation}{Equations (#2#1#3)}{ and~(#2#1#3)}{, (#2#1#3)}{ and~(#2#1#3)} 

\newcommand{\major}[1]{#1}

\begin{document}

\title{Active suspensions have non-monotonic flow curves and multiple mechanical equilibria}

\author{Aurore Loisy}
\author{Jens Eggers}
\author{Tanniemola B. Liverpool}
\affiliation{School of Mathematics, University of Bristol - Bristol BS8 1TW, UK}

\date{\today}

\begin{abstract}
We point out unconventional mechanical properties of confined active fluids, such as bacterial suspensions, under shear.
Using a minimal model of an active liquid crystal with no free parameters, we predict the existence of a window of bacteria concentration for which a suspension of \textit{E.~Coli} effectively behaves, at steady-state, as a negative viscosity fluid and reach quantitative agreement with experimental measurements.
Our theoretical analysis further shows that a negative apparent viscosity is due to a non-monotonic local velocity profile, and is associated with a non-monotonic stress vs. strain rate flow curve. 
This implies that fixed stress and fixed strain rate ensembles are not equivalent for active fluids.
\end{abstract}


\maketitle

Active suspensions, such as swarms of bacteria and the cytoskeleton of living cells, consist of interacting self-driven particles that individually consume energy and collectively generate motion and mechanical stresses in the bulk \cite{Toner2005,Ramaswamy2010,Marchetti2013,Saintillan2013,Prost2015}.
Due to the orientable nature of their constituents, active suspensions can exhibit liquid crystalline order and have been modeled as active liquid crystals (LCs) \cite{Ramaswamy2010,Marchetti2013,Prost2015}. 
An astonishing property of confined active LCs is their ability to spontaneously flow in the absence of any mechanical forcing \cite{Voituriez2005,Voituriez2006,Marenduzzo2007a,Marenduzzo2007b,Giomi2008,Edwards2009,Furthauer2012}. 
It is therefore expected that active LCs exhibit, under external driving, unusual mechanical properties.

\begin{figure}
	\centering
	\includegraphics[width=0.99\linewidth]{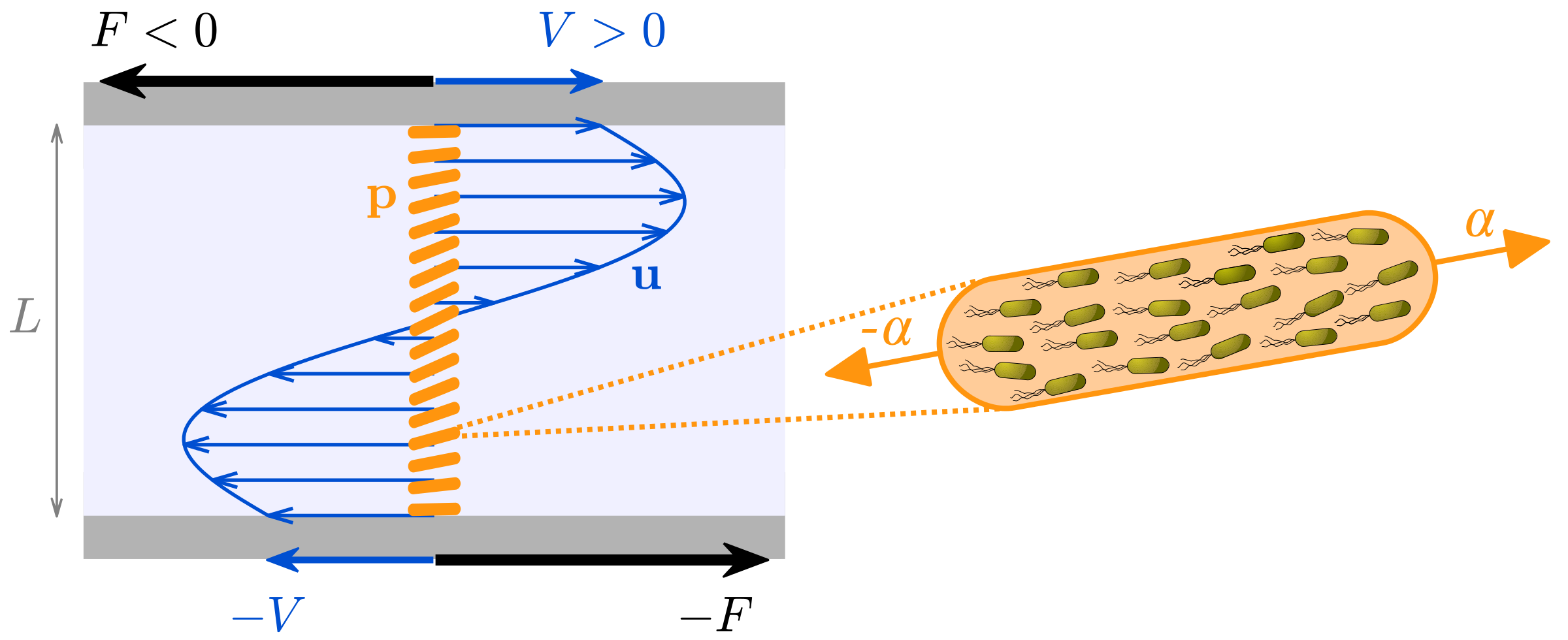}
	\caption{
	A suspension containing orientable active units such as bacteria is sheared between two plates of area $A$ separated by a distance $L$. A force $F$ is exerted on the plates which move at a velocity $V$. 
	The apparent viscosity, defined as $\eta_{\mathrm{app}}=\sigma/\dot{\gamma}$ where $\sigma=F/A$ is the macroscopic shear stress and $\dot{\gamma}=2V/L$ is the macroscopic shear strain rate, can become negative under certain conditions.
	When $\eta_{\mathrm{app}} < 0$, the velocity profile is non-monotonic with local velocity gradients at the walls opposing the applied macroscopic velocity gradient.
	\label{fig:negative_viscosity_illustration}
	}
\end{figure}

The most commonly sought mechanical property of a complex fluid is the apparent (shear) viscosity, which expresses the macroscopic fluid's resistance to flow and can be defined through the idealized configuration depicted in \cref{fig:negative_viscosity_illustration}. 
A uniform shear is applied to the fluid by confining it between two moving parallel plates, and the apparent viscosity of the fluid is $\eta_{\mathrm{app}}=\sigma/\dot{\gamma}$ where $\sigma$ is the \emph{macroscopic} shear stress and $\dot{\gamma}$ is the \emph{macroscopic} shear strain rate. 
We emphasize that $\sigma$ and $\dot{\gamma}$ are spatially averaged over the gap, they are not equivalent to their local counterparts which may not be uniform.

There is now ample evidence that the apparent viscosity of suspensions can be reduced by activity \cite{Hatwalne2004,Liverpool2006,Haines2009,Saintillan2010,Saintillan2010a,Ryan2011,Slomka2017b,Sokolov2009,Gachelin2013,Lopez2015,Saintillan2013}.
But only recently has it been taken seriously that this phenomenon can continue until even a negative apparent viscosity is achieved. 
Using a highly sensitive rheometer, Lopez et al. \cite{Lopez2015} were able to measure zero and possibly negative values of the apparent viscosity in a suspension of \textit{E.~Coli} at steady-state, thereby demonstrating that microscopic bacterial activity can be converted into macroscopic useful mechanical power.
Several theories have been proposed since to rationalize $\eta_{\mathrm{app}}\le0$, either based on kinetic models \cite{Nambiar2017,Takatori2017} or on a generalized Navier-Stokes equation \cite{Slomka2017b}, yet comparison with \cite{Lopez2015} is, at best, qualitative.

\begin{figure}
	\begin{flushleft}
		(a) \hspace{0.43\linewidth} (b)
	\end{flushleft}
	\vspace{-1em}
	\centering
	\includegraphics[width=0.49\linewidth]{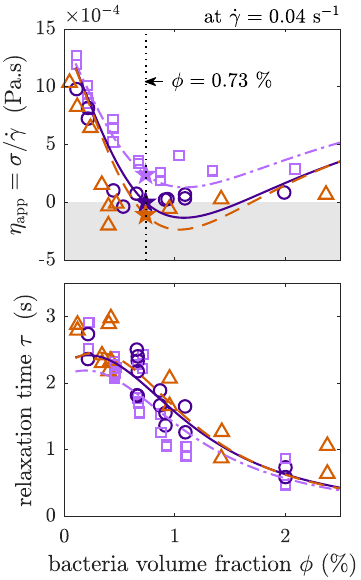}
	\includegraphics[width=0.49\linewidth]{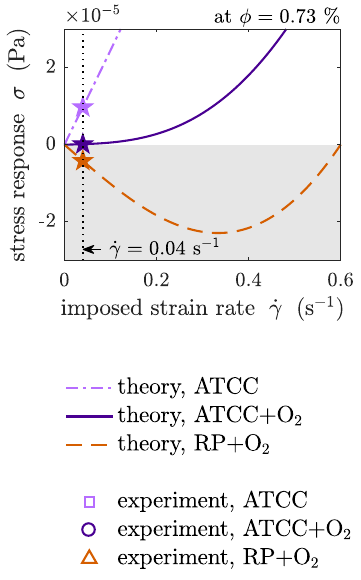}
	\caption{
	(a) Effect of bacteria volume fraction on the steady-state apparent viscosity (top) and on the relaxation time (bottom) of \textit{E.~Coli} suspensions subjected to a step change of strain rate from 0 to $\dot{\gamma}=0.04$~s$^{-1}$: present theory with no free parameters (lines) and experiments (open symbols) \cite{Lopez2015}.
	This behavior was not explained by prior models \cite{Nambiar2017,Takatori2017}.
	(b) Theoretical \major{macroscopic} flow curves $\sigma(\dot{\gamma})$ for \textit{E.~Coli} suspensions with $\eta_{\mathrm{app}}>0$, $\eta_{\mathrm{app}}=0$ and $\eta_{\mathrm{app}}<0$. \major{Suspensions with $\eta_{\mathrm{app}}<0$ have non-monotonic $\sigma(\dot{\gamma})$.}
	In (a-b), the different colors, symbols and line styles corresponds to different bacterial strains (ATCC and RP) and oxygen levels, and the stars mark corresponding state points.
	\label{fig:theory_vs_exp}
	}
\end{figure}

In this Letter, we show that a minimal model of an active LC can\major{, without shear bands \cite{Olmsted2008,Cates2008,Fielding2011}}, predict \emph{quantitatively} the transition to $\eta_{\mathrm{app}}\le0$ reported by \cite{Lopez2015}. Our model, which has no free parameters, explains both the observed non-monotonic evolution of $\eta_{\mathrm{app}}$ and the decreasing response time of the system with increasing bacterial concentration (\cref{fig:theory_vs_exp}a).
Importantly $\eta_{\mathrm{app}}<0$ is due to a non-monotonic velocity profile (\cref{fig:negative_viscosity_illustration}), the local viscosity being always positive (essentially that of the solvent).
We further show that the steady state with $\eta_{\mathrm{app}}<0$ reached by \cite{Lopez2015} is a structurally stable mechanical equilibrium, and is associated with a non-monotonic \major{macroscopic} flow curve $\sigma(\dot{\gamma})$ (\cref{fig:theory_vs_exp}b).
As a consequence, if the suspension were to be sheared by tuning the stress rather than the strain rate, experiments would yield different results: $\sigma(\dot{\gamma})\not\equiv\dot{\gamma}(\sigma)$. This also implies history dependent mechanics.

The hydrodynamic theory of active matter provides a now well-accepted continuum description of active LCs in terms of a reduced number of slowly-varying fields \cite{Marchetti2013,deGennes1993book}. 
The local coarse-grained orientation of the particles is represented by the polarization vector.
The magnitude of the polarization vector is a fast variable, therefore we take it to be constant on long timescales and assume without loss of generality that the polarization is a unit vector $\v{p}$ (the effect of its magnitude being absorbed in the phenomenological coefficients).
The other relevant slow variables are the fluid velocity $\v{u}$ and the particle number density. 
For simplicity we shall assume nematic symmetry and homogeneous density, this hypothesis is relaxed and shown to be unimportant in Supplemental Material.

The evolution of the director field, in a globally ordered or isotropic phase (bearing in mind that local alignment is always present), is governed by
\begin{equation}
	\left( \partial_t + u_j \partial_j \right) p_i + \Omega_{ij} p_j = \lambda E_{ij} p_j + \Gamma h_i
\end{equation}
where $E_{ij}=(\partial_i u_j + \partial_j u_i)/2$, $\Omega_{ij}=(\partial_i u_j - \partial_j u_i)/2$, $\lambda$ is the flow alignment parameter which determines the response of $\v{p}$ to simple unbounded shear ($\abs{\lambda}>1$ for alignment, $\abs{\lambda}<1$ for tumble), $1/\Gamma$ is the rotational viscosity, and $h_i=K\nabla^2 p_i+h_\parallel^0 p_i$ is the molecular field where $K$ 
\major{assigns a cost to distortion ($K$ is an \emph{effective} Frank elasticity and both $K,\lambda$ could include nonequilibrium contributions)}
and $h_\parallel^0$ is a Lagrange multiplier enforcing $\norm{\v{p}}=1$. 
The fluid is assumed incompressible ($\div\v{u} = 0$) and the flow field obeys, upon neglecting fluid inertia, the Stokes flow equation $\div \m{\sigma} = 0$ with
\begin{equation}
	\sigma_{ij} = 2 \eta E_{ij} - \Pi \delta_{ij} - \frac{\lambda+1}{2} p_i h_j - \frac{\lambda-1}{2} p_j h_i - \alpha p_i p_j
\end{equation}
where $\eta$ is the bulk fluid viscosity ($\eta>0$), $\Pi$ is the bulk pressure, and $\alpha$ is the activity coefficient.
This coefficient is related to the active stresses in a suspension of particles modeled as force dipoles: the magnitude of $\alpha$ is proportional to the strength of the force pair and the sign of $\alpha$ depends on whether the induced flow is extensile ($\alpha>0$) or contractile ($\alpha<0$).
Our geometry is a two-dimensional slab of thickness $L$ (\cref{fig:negative_viscosity_illustration}) with translational invariance in the direction parallel to the walls. We use no-slip and parallel anchoring as boundary conditions.
The fluid is subject to a macroscopic shear rate $\dot{\gamma}=2V/L$ and the shear stress (simply denoted $\sigma$) is uniform across the film. 

\begin{figure*}
	\begin{flushleft}
		(a)
	\end{flushleft}
	\vspace{-1.2em}
	\centering
	\includegraphics[width=0.99\linewidth]{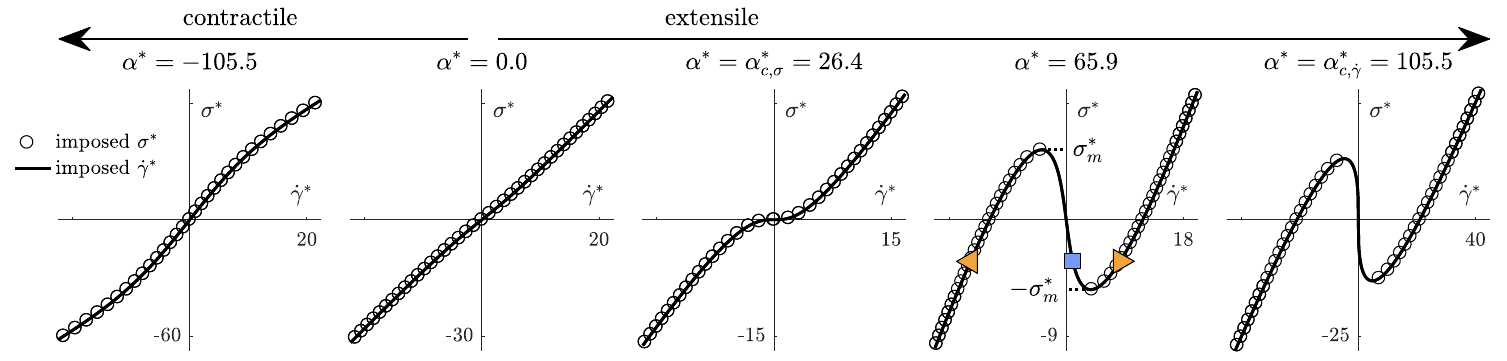}
	\begin{flushleft}
		(b) \hspace{0.37\linewidth} (c) \hspace{0.37\linewidth} (d)
	\end{flushleft}
	\vspace{-1.2em}
	\centering
	\includegraphics[width=0.4\linewidth]{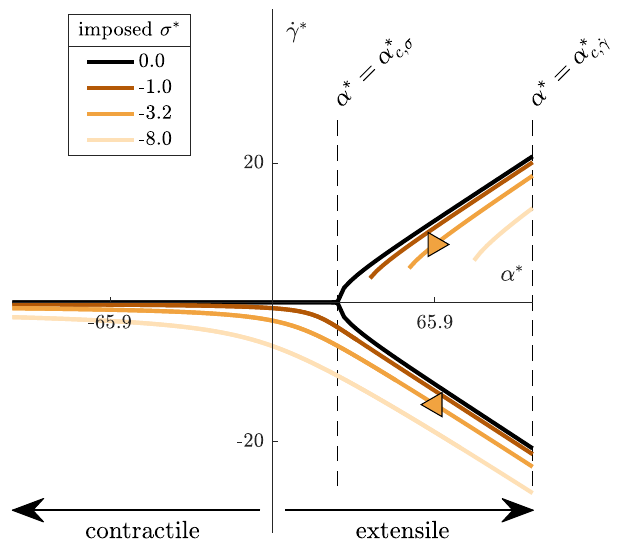}
	\includegraphics[width=0.4\linewidth]{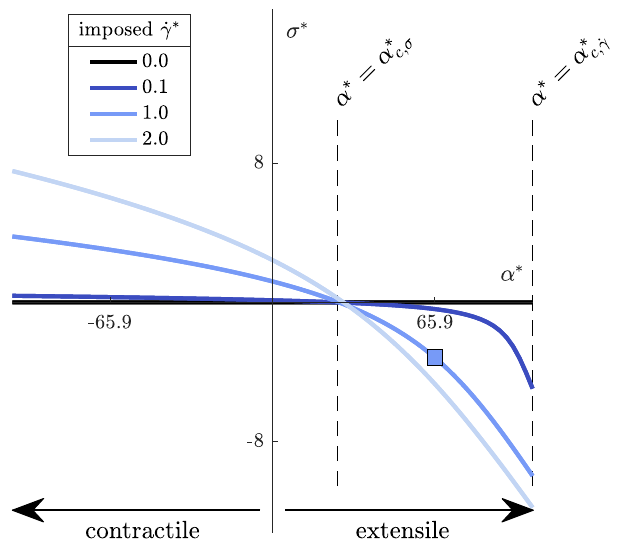}
	\includegraphics[width=0.18\linewidth]{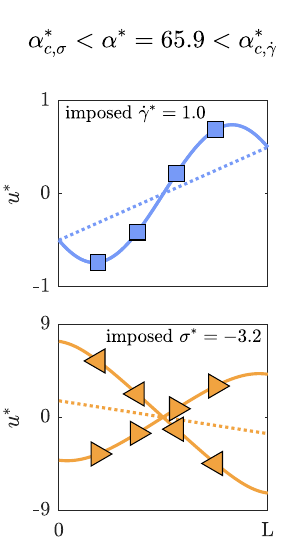}
	\caption{
	Steady-state \major{macroscopic} rheology of a \emph{weakly} sheared active suspension of flow-aligning particles ($\lambda=1.9$).
	(a) flow curves in stress-controlled (circles) and strain-rate-controlled (lines) conditions for increasing $\alpha$ (from left to right);
	(b-c) bifurcation diagrams in stress-controlled (b) and strain-rate-controlled (c) conditions.
	The symbols show the correspondence between the flow curves and the bifurcation diagrams, associated velocity profiles are provided in (d) (for comparison, dotted lines show the behavior of a Newtonian passive fluid).
	Only stable solutions are shown.
	In the absence of applied shear, a pitchfork bifurcation \major{(associated to the spontaneous flow transition found by \cite{Voituriez2005})} occurs at $\alpha=\alpha_{c,\sigma}$ for $\sigma=0$ and at $\alpha=\alpha_{c,\dot{\gamma}}$ for $\dot{\gamma}=0$.
	These bifurcations correspond to the singularities ($\d\dot{\gamma}/\d\sigma|_{\sigma=0} = \infty$ and $\d\sigma/\d\dot{\gamma}|_{\dot{\gamma}=0} = -\infty$, respectively) in the flow curves.
	The stars denote dimensionless quantities: $\alpha^*=\alpha L^2/K$, $\sigma^*=\sigma L^2/K$, $\dot{\gamma}^*=\dot{\gamma} L^2/(\Gamma K)$, and $u^*= u L/(\Gamma K)$. 
	These results are qualitatively independent of $\lambda$. 
	\label{fig:flow_curves_bifurcation}
	}
\end{figure*}

To determine the apparent viscosity of a fluid we have the choice of two ensembles: prescribed $\dot{\gamma}$ or prescribed $\sigma$. 
In passive fluids, both ensembles are equivalent: the steady shear response is characterized uniquely by $\sigma(\dot{\gamma})$ or by $\dot{\gamma}(\sigma)$.
We emphasize here that this is not generally true in active fluids.
This peculiar property may be seen in the limiting cases $\dot{\gamma}=0$ or $\sigma=0$. 
In the absence of external mechanical forcing, a confined quasi-one-dimensional active LC exhibits a spontaneous transition from a homogeneous immobile state to an inhomogeneous flowing state at a critical value of the activity coefficient $\alpha$ \cite{Voituriez2005}.
Crucially, the spontaneous flow transition depends on the mechanical constraint: it occurs at $\alpha_{c,\sigma}$ for $\sigma=0$ but at a larger value $\alpha_{c,\dot{\gamma}}=4\alpha_{c,\sigma}$ for $\dot{\gamma}=0$, where $\alpha_{c,\sigma} =  2 \upi^2 \eta \Gamma K L^{-2} (1 + \xi)/(\lambda-1)$ 
with $\xi=(\lambda-1)^2/(4 \eta \Gamma)$ \cite{Voituriez2005}.
The immediate consequence is that $\sigma(\dot{\gamma}=0)=0$ whereas $\dot{\gamma}(\sigma=0) \neq 0$ for $\alpha_{c,\sigma}<\alpha<\alpha_{c,\dot{\gamma}}$ \footnote{For $\alpha_{c,\sigma}>0$, otherwise the order relations are reversed}. 
Note that the spontaneous flow transition is only seen in flow-aligning extensile systems ($\lambda>1$, $\alpha>0$) or flow-tumbling contractile systems ($0<\lambda<1$, $\alpha<0$) \cite{Marenduzzo2007b,Edwards2009} \footnote{For simplicity and without loss of generality we restrict to $\lambda>0$}.
 
By studying numerically the same active LC in the presence of external shear (see Supplemental Material for methodology details), we found that this system can also undergo a bifurcation under \emph{weak} applied shear and exhibit two asymmetric stable branches beyond $\alpha_{c,\sigma}$ for imposed $\sigma$ (\cref{fig:flow_curves_bifurcation}b) or $\alpha_{c,\dot{\gamma}}$ for imposed $\dot{\gamma}$ (not shown: as such a high level of activity is irrelevant to experiments, results are deferred to Supplemental Material).
This results in the non-equivalence of these two ensembles and in the unconventional flow curves $\sigma(\dot{\gamma})$ and $\dot{\gamma}(\sigma)$ displayed in \cref{fig:flow_curves_bifurcation}a.

For $\alpha<\alpha_{c,\sigma}$, the flow curves are identical and increase monotonically, as in passive fluids.
When $\alpha=\alpha_{c,\sigma}$, $\dot{\gamma}(\sigma)$ has a vertical tangent at zero, which corresponds to the bifurcation from a single steady state to bistability for imposed $\sigma$.
For $\alpha_{c,\sigma}<\alpha<\alpha_{c,\dot{\gamma}}$, $\sigma(\dot{\gamma})$ exhibits local extrema $\pm \sigma_m$ connected by a negative slope $\d\sigma/\d\dot{\gamma} < 0$. The associated solutions are unique and stable when $\dot{\gamma}$ is imposed (blue square), but are unstable (not shown) when $\sigma$ is imposed. Instead $\dot{\gamma}(\sigma)$ has two stable non-equivalent solutions (leftward and rightward orange triangles) in the range $[-\sigma_m,+\sigma_m]$. 
While these features were identified separately by \cite{Giomi2010} for the former and by \cite{Furthauer2012} for the latter, here we provide the conceptual framework that shows they are simply different sides of the same coin. 
When $\alpha=\alpha_{c,\dot{\gamma}}$, $\sigma(\dot{\gamma})$ admits a vertical tangent at zero, which corresponds to the bifurcation for imposed $\dot{\gamma}$.
For $\alpha>\alpha_{c,\dot{\gamma}}$, $\sigma(\dot{\gamma})$ exhibits discontinuous branches which shape depends on $\lambda$ as explained in Supplemental Material.

A related, and striking, property of weakly sheared active films is the existence, for $\alpha_{c,\sigma}<\alpha<\alpha_{c,\dot{\gamma}}$, of a
\emph{structurally stable} mechanical equilibrium with $\eta_{\mathrm{app}}<0$ if (and only if) $\dot{\gamma}$ is imposed (blue square in \cref{fig:flow_curves_bifurcation}). 
It is accommodated through a sinusoidal modulation of the linear velocity profile (\cref{fig:flow_curves_bifurcation}d, blue squares) rather than through well-defined shear bands \cite{Cates2008,Fielding2011}, and corresponds to the regime reached in the experiments of \cite{Lopez2015}.
If one would tune the applied stress instead, $\dot{\gamma}(\sigma)$ would exhibit hysteresis and at $\sigma=0$, the plates would move.

\begin{figure}
	\includegraphics[width=0.99\linewidth]{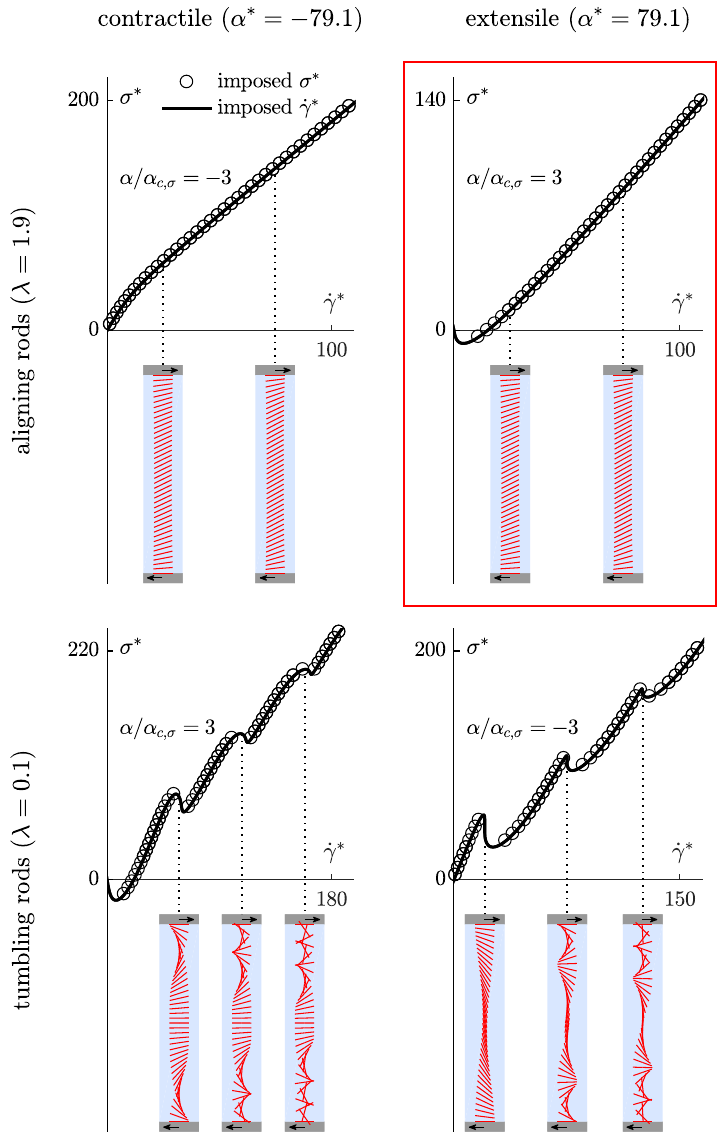}
	\caption{
	Steady-state \major{macroscopic} flow curves for various types of active suspensions (insets: steady-state profiles of $\v{p}$). 
	Bacteria suspensions of \cite{Lopez2015} follow the scenario displayed in the framed panel.
	The stars denote dimensionless quantities, see \cref{fig:flow_curves_bifurcation}.
	\label{fig:flow_curves_wiggles}
	}
\end{figure}

When an active film is \emph{strongly} sheared, its rheology can be either Newtonian or strongly nonlinear depending on the magnitude of $\lambda$ (\cref{fig:flow_curves_wiggles}).
For $\abs{\lambda}>1$, the particles respond to shear by aligning at a well-defined angle with respect to the flow direction. Suspensions in this flow-aligning regime behave as Newtonian fluids under strong shear.
For $\abs{\lambda}<1$, the particles rotate and form rolls within the gap as applied shear increases. The flow curves of suspensions in this flow-tumbling regime are characterized by multiple regions of locally reduced apparent viscosity which coincide with the completion of a half turn by the director at the center of the film. 
While these nonlinearities are also present in passive systems, non-monotonic flow curves with multiple local extrema or discontinuities as in \cref{fig:flow_curves_wiggles} are peculiar to active systems.
The sign of $\lambda$, typically (but not necessarily) related to the particle shape, plays no role here (see Supplemental Material for $\lambda<0$).

These theoretical predictions have been obtained from generic equations which apply to a broad class of active LCs \cite{Marchetti2013}.
The price to pay for this generality is the introduction of several system-dependent coefficients.
In the following we will show how these phenomenological coefficients can be extracted from shear experiments to allow testable quantitative predictions.

For that purpose we consider the \emph{transient} response of an active film to applied shear and restrict our analysis to small $\sigma$ or $\dot{\gamma}$.
By solving the linearized governing equations (see Supplemental Material) we find that the strain rate response to a step of stress from $0$ to $\sigma$, valid for $\alpha/\alpha_{c,\sigma}<1$, is
\begin{subequations}
\label[equation]{eq:linear_solution_sigma}
\begin{align}
	\frac{\sigma}{\dot{\gamma}(t)} & = \eta \bigg[ \frac{\eta}{\eta_{\mathrm{app}}} - \Big(\frac{\eta}{\eta_{\mathrm{app}}} - 1 \Big) \exp \Big( -\frac{t}{\tau_{\sigma}} \Big) \bigg]^{-1} \\
\intertext{with the relaxation time}
	\frac{1}{\tau_{\sigma}} & = \frac{\upi^2 \Gamma K}{L^2} \bigg[ 1 + \xi - \dfrac{\alpha}{\alpha_{c,\sigma}} (1+\xi)  \bigg]
\end{align}
\end{subequations}
whereas the stress response to a step of strain rate from $0$ to $\dot{\gamma}$, valid for $\alpha/\alpha_{c,\dot{\gamma}}<1$, reads
\begin{subequations}
\label[equation]{eq:linear_solution_gammadot}
\begin{align}
	\frac{\sigma(t)}{\dot{\gamma}} & = \eta \bigg[ \frac{\eta_{\mathrm{app}}}{\eta} - \Big(\frac{\eta_{\mathrm{app}}}{\eta} - 1 \Big) \exp \Big( -\frac{t}{\tau_{\dot{\gamma}}} \Big) \bigg] \\
\intertext{with the (different) relaxation time}
	\frac{1}{\tau_{\dot{\gamma}}} & = \frac{\upi^2 \Gamma K}{L^2} \bigg[ 1 + \Big( \xi - \dfrac{\alpha}{\alpha_{c,\sigma}} \left( 1 + \xi \right) \Big) \Big( 1 - \frac{8}{\upi^2} \Big) \bigg].
\end{align}
\end{subequations}
The signal of $\sigma/\dot{\gamma}$ initially jumps to $\eta$ and then relaxes to $\eta_{\mathrm{app}}$ given by 
\begin{equation}
	\frac{\eta_{\mathrm{app}}}{\eta} = \frac{1 - \dfrac{\alpha}{\alpha_{c,\sigma}}}{ 1 - \dfrac{8}{\upi^2} \dfrac{\xi}{\xi + 1} - \left(1 - \dfrac{8}{\upi^2}\right) \dfrac{\alpha}{\alpha_{c,\sigma}}}.
	\label{eq:linear_apparent_viscosity}
\end{equation}
\Cref{eq:linear_apparent_viscosity} shows that an active suspension behaves, at low applied strain rate, as a \emph{Newtonian} fluid with $\eta_{\mathrm{app}}\leqslant0$ when $\alpha/\alpha_{c,\sigma}\geqslant1$ (as also visible in \cref{fig:flow_curves_bifurcation}c).

The four independent (groups of) parameters $\eta$, $\alpha(\lambda-1)$, $\Gamma K$ and $\xi$ control the suspension behavior. They can all be identified from rheological time traces using \cref{eq:linear_solution_sigma,eq:linear_solution_gammadot,eq:linear_apparent_viscosity}. 
We applied this to the \textit{E.~Coli} suspensions studied by \cite{Lopez2015}: using their experimental time signals of $\sigma(t)/\dot{\gamma}$, we inferred the values of the phenomenological parameters describing their suspensions for bacteria volume fractions $\phi$ ranging from 0.11~\% to 2.4~\% (see Supplemental Material for more details about our procedure and the inferred parameters).

We obtained $\xi\approx0$, $\eta\approx\eta_s$ with $\eta_s$ the solvent viscosity, and $\alpha (\lambda-1)$ and $\Gamma K$ were found to be positive increasing functions of $\phi$.
Since \textit{E.~Coli} are extensile swimmers or ``pushers'' ($\alpha>0$) \cite{Drescher2011}, it follows that $\lambda>1$, i.e., \textit{E.~Coli} behave effectively as flow-aligning rod-like particles. 
The dependences on $\phi$ are consistent with the expectations that (i) $\alpha=\alpha_1 \phi$ for a dilute suspension of self-propelled swimmers and (ii) $K=K_0 + K_1 (\phi-\phi_c)^2$ for a nematic phase \footnote{Below the transition, one would expect instead $K \propto \phi^2$, which is incompatible with the data.} \cite{Tjipto-Margo1992}, where $K_{0,1}>0$ and where $\phi_c$ is the critical volume fraction (smaller than 0.1 \%) associated to the isotropic-nematic transition \major{(as $K$ includes nonequilibrium contributions, the constants may differ significantly from their equilibrium counterparts)}.
Plugging semi-analytical expressions of $\alpha(\phi)$ and $K(\phi)$ into the governing equations yields, without any adjustable parameter, the theoretical curves presented against the experimental data in \cref{fig:theory_vs_exp}a.

We conclude that a minimal model of an active LC is sufficient to account quantitatively for all experimental observations (\cref{fig:theory_vs_exp}), notably the emergence of a ``superfluid-like'' regime with $\eta_{\mathrm{app}}\lesssim0$ \major{(which corresponds to $\alpha/\alpha_{c,\sigma}\gtrsim1$) over a window of bacteria volume fractions (around 1 \%).}
The non-monotonic evolution of $\eta_{\mathrm{app}}$ with $\phi$ results from \major{the competition between activity $\alpha$ and effective stiffness $K$}.
Incidentally, the predicted increase of $\eta_{\mathrm{app}}$ at higher volume fractions, although not clearly seen in the experiments of \cite{Lopez2015}, resemble that reported by \cite{Sokolov2009} for a suspension of \textit{Bacillus subtilis}, an analogous type of rod-like pushers. 
\major{Finally, the increase of $K$ with $\phi$ explains the reduction of the response time $\tau$ with increasing $\phi$.}

Geometrical confinement provides a powerful way to control active suspensions \cite{Wioland2013,Wioland2016,Wu2017,Theillard2017}. 
The existence of non-monotonic flow curves suggest new control mechanisms, that could find direct application in bacterial energy harvesting \cite{Sokolov2010,DiLeonardo2010}.
Our results also provide strong support for models of biologically active suspensions as ``living liquid crystals'' \cite{Ramaswamy2010,Marchetti2013,Prost2015} and open the way to a truly quantitative characterization of these systems.

\begin{acknowledgments}
Part of this work was funded by a Leverhulme Trust Research Project Grant RPG-2016-147. We are grateful to \'E. Cl\'ement and H. Auradou for sharing experimental data. We thank D. Cortese for helpful discussions.  
TBL acknowledges support of BrisSynBio, a BBSRC/EPSRC Advanced Synthetic Biology Research Centre (grant number BB/L01386X/1).
\end{acknowledgments}


%

\end{document}